# Reliability Study of Power Harvesting System from Sea Waves with Piezoelectric Patches

Hadi Mirab[1], Vahid Jahangiri[2]*, Mir Mohammad Ettefagh[1], Reza Fathi[1]
[1] Dept. of Mechanical Engineering, University of Tabriz, Tabriz, Iran.
[2] Dept. of Civil Engineering, Louisiana State University, Baton Rouge, USA. Vjahan1@lsu.edu

**Abstract**
Conversion of sea waves mechanical energies into electrical form of energy by means of piezoelectric materials is considered as one of the most recent methods for powering low-power electronic devices at sea. In this paper, power harvesting from sea waves by consideration of JONSWAP wave theory is investigated and the uncertainties of sea waves are studied for the first time. For this purpose, a vertical beam fixed to the seabed which the piezoelectric patches are attached to it, is considered as energy harvester and is modeled and simulated by MATLAB software. The generated power is computed by calculating the beam vibration response and the effect of piezoelectric patches on the generated power is studied by statistical analysis. Furthermore, reliability of the energy harvesting system is investigated as the possibility of failure based on violation criteria. It is resulted that the probability of failure increases by increasing the power.

## 1. Introduction

Implementation of wireless communication and low-power electricity sensors provide rich information for different industries such as alarming devices, structural health monitoring and marine industry. The low power electric devices can be used as energy supplies for wireless sensors, instead of batteries which have disadvantages in comparison with the low power electric devices. The increasing demand of wireless communication devices in various industries has led researchers to investigate on converting ambient vibration into electrical energy by implementing piezoelectric, electromagnetic and electrostatic material. However, utilizing piezoelectric material has been taken into consideration more than the other materials due to the benefits of these kinds of materials . Harvesting energy by using piezoelectric converters has been a subject of many types of researches. Anton and Sodano [1]–[3] reviewed recent literature in the field of energy harvesting. Smits et al. [4], [5] derived the constituent equations of a heterogeneous piezoelectric bender under different boundary conditions, where these boundary conditions consists a mechanical moment at the end, a force which is applied to the tip, and a uniform load which is applied over the entire length of the bender. They also discussed the electromechanical characteristics of the mentioned bender. Zhang and Lin [6] summarized and illustrated the behavior of several types of ocean wave energy converters using piezoelectric materials. They also designed a novel wave energy converter by implementation of piezoelectric material. Tanaka et al. [7] discussed forced vibration experiments on flexible piezoelectric devices which operates in both air and water environments. They also manufactured and tested several devices in different operating conditions. Finally, they compared the experimental and simulation results. Viet et al. [8] developed a floating kind of energy harvester system using the piezoelectric materials to scavenge the energy from the motion of waves. They concluded that the harvested power increases by an increase in the ocean wave amplitude and a decrease in the ocean wave period. Taylor et al. [9] designed a device made of a flexible polyvinylidene fluoride, which converts the mechanical flow energy, available in rivers to electrical power. Wu et al. [10] developed a more efficient energy harvesting method by piezoelectric materials. They also proposed a numerical model to compute the generated electrical power. It was resulted that higher electrical power can be achieved when the harvester has a thinner and longer floater. Zurkinden et al. [11] designed some similar wave energy converter devices from piezoelectric material, which are subjected to the wave force at a characteristic wave frequency to scavenge energy from the ocean surface waves. Xie et al. [12] designed a piezoelectric coupled plate structures fixed on the seabed and a base structure in the case of harvesting the ocean wave energy. It was shown that the produced electrical energy increases by increasing the length of cantilevers and the wave height and by decreasing the distance of the ocean surface to the cantilevers. One of the most recent researches on energy harvesting from sea waves is done by Wang et al. [13]. They studied the effect of energy harvester system and wave parameters on generated electrical power produced from piezoelectric energy harvester system according to the Airy linear wave theory. Mirab et al. [14] studied energy harvesting system with considering both JONSWAP and Airy wave theory. Moreover, annealing algorithm was used to optimize the parameters of energy harvesting system.

In this paper, an energy harvesting system is subjected to sea wave which is modeled by accurate JONSWAP random wave theory. Unlike the Airy wave theory, JONSWAP wave model contains random features of sea waves. For this purpose, an energy harvester which is jointed to the seabed according to irregular JONSWAP wave model is considered. The random phenomenon is made because of irregularities which can be created due to uncertainties in characteristics of the system or force excitations. It should be noted that the present study focuses on both randomnesses of the system excitations and uncertainties in generated electrical power. After modeling and simulating the system in MATLAB, the exact vibration response of the beam is derived by numerical methods. By knowing the displacements of the beam, the generated power is calculated and the effects of uncertainties related to the piezoelectric patches on the generated power are investigated. In addition, statistical analysis and probability of failure are studied. Finally, it was shown that the probability of failure increases by increasing the power.

## 2. Modeling of energy harvester system by considering JONSWAP wave theory

In this section, the energy harvesting system is modeled and its equations of motion are derived. In addition, JONSWAP wave theory is explained and defined. Finally, the generated electrical power is calculated.

### 2.1. Vibration equation of beam under excitation of sea waves

A cantilever beam with length of $l$, and piezoelectric patches that are attached to the beam, is the most current device in scavenging energy from vibration energy which is shown in Fig. 1.

The beam is fixed to the seabed in depth of $d$, and the beam starts to vibrate when it is subjected to sea waves and this causes dynamical strain in piezoelectric layers which produces electrical power. Transverse motion of beam can be expressed as Eq. (1):

$$EI\frac{\partial^4 w(z,t)}{\partial z^4} + m'g\frac{\partial^2 w(z,t)}{\partial z^2} + \rho A\frac{\partial^2 w(z,t)}{\partial t^2} = f_H(t,z) \qquad (1)$$

Where $m'$ is the mass per unit length of the beam at each position of $z$, $EI$ and $A$ are the bending rigidity and cross section area of the beam respectively, $\rho$ is the density, $w(z,t)$ is the longitudinal displacement of the beam at position $z$ and $f_H(t,z)$ is the force resulted from sea waves.

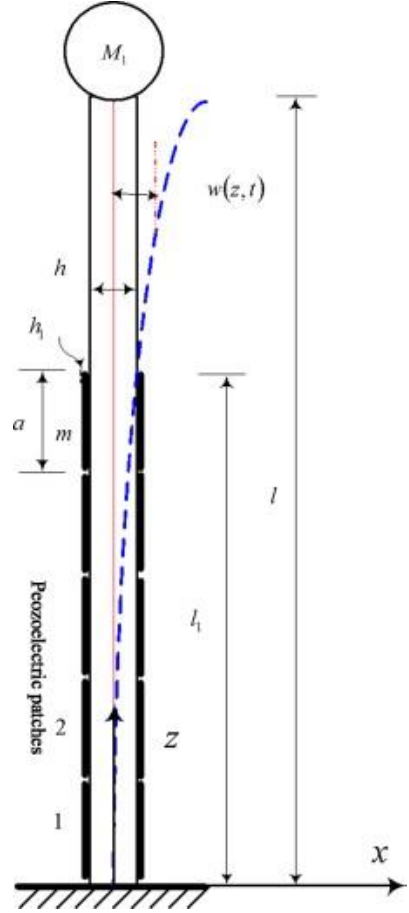

**Fig.1** Set up of the piezoelectric energy harvester (11)

The horizontal wave force is determined from Morison equation as follows [15]:

$$f_H(t,z) = \frac{1}{2}c_D\rho_w b\left(u_x - \frac{\partial w}{\partial t}\right)\left|u_x - \frac{\partial w}{\partial t}\right| + c_M\rho_w bh a_x - c_m\rho_w bh\frac{\partial^2 w}{\partial t^2} \qquad (2)$$

where $c_D$, $c_M$ and $c_m$ are the coefficient of the drag and inertia forces of the beam and the added mass respectively, $\rho_w$ is the material density of the beam, $h$ and $b$ are the thickness and width of beam respectively. Also, $u_x$ and $a_x$ are the longitudinal velocity and acceleration of the water particles in the sea respectively. In this paper, in order to determine velocity and acceleration presented in Morrison equation, JONSWAP irregular wave model is considered where it can be defined in [16].

**2.2. Derivation of vibrational response**

By having the value of the energy harvester force, result of Eq. (1) can be calculated by applying mode summation and variable separation method as follows:

$$w(z,t) = \sum_{i=1}^{3} W_i(z) q_i(t) \tag{3}$$

where $W_i(z)$, is the shape mode function and $q_i(t)$ is the response in the time domain. Mode function for transverse vibration of the beam is expressed as:

$$W_i(z) = c_1 \cosh s_1 z + c_2 \sinh s_1 z + c_3 \cos s_2 z + c_4 \sin s_2 z \tag{4}$$

where $s_{1,2}$ are determined from Eq. (5). The boundary conditions are introduced as: displacement and rotation are zero at the fixed point. Also, bending moment of the beam is zero at the free point. In addition, the shear force of the beam is equal to inertial force of the point mass. According to these boundary conditions $c_{1,2,3,4}$ coefficients are calculated.

$$s_{1,2} = \sqrt{\pm \frac{m'g}{2EI} + \sqrt{\frac{[m'g]^2}{4(EI)^2} + \frac{\omega^2(k_3 + \rho A)}{EI}}} \tag{5}$$

**2.3. Response in time domain**

In the case of deriving responses in time domain, response of the beam is expanded according to Eq. (3) and also mode shape orthogonality is used. Response in the time domain and vibration response of energy harvester are calculated by solving these equations numerically. By substituting Eq .(3) in Eq. (1), the new equation will be as follows:

$$EI \sum_{i=1}^{N} W_i^4(z) q_i(t) + M_1 g \sum_{i=1}^{N} W_i''(z) q_i(t) + (k_3 + \rho A) \sum_{i=1}^{N} W_i(z) \ddot{q}_i(t) = f_H(t,z) \tag{6}$$

By multiplying two sides of Eq. (6) in $W_j(z)$ or by integrating over the length of the beam from $-d$ to $0$, Eq. (7) is derived as follows:

$$EI \sum_{i=1}^{N} \int_{-d}^{0} W_i^{(4)}(z) dz q_i(t) + M_1 g \sum_{i=1}^{N} \int_{-d}^{0} W_i''(z) dz q_i(t) + M_1 g \sum_{i=1}^{N} \int_{-d}^{0} W_i''(z) W_j(z) dz q_i(t)$$
$$+ (k_3 + \rho A) \sum_{i=1}^{N} \int_{-d}^{0} W_i(z) W_j(z) dz \ddot{q}_i(t) = \int_{-d}^{0} f_H(t,z) W_j(z) dz \tag{7}$$

By expanding the latest equation, Eq .(8) is determined:

$$[M]\{\ddot{q}(t)\} + [K]\{q(t)\} = \{Q\} \tag{8}$$

where $[M]$ is mass matrix, $[K]$ is the stiffness matrix, $\{Q\}$ is generalized force, $\{q(t)\}$ is generalized coordinate and $\{\ddot{q}(t)\}$ is second derivative of generalized coordinate whereas the elements of these matrices are shown in Eq. (9) to Eq. (11) respectively.

$$M_{ij} = (k_3 + \rho A) \int_{-d}^{0} W_i(z) W_j(z) dz \tag{9}$$

$$K_{ij} = EI \int_{-d}^{0} W_j^{(4)}(z) W_i(z) dz + M_1 g \int_{-d}^{0} W_j''(z) W_i(z) dz \tag{10}$$

$$Q_i = \int_{-d}^{0} f_H(t,z) W_i(z) dz \tag{11}$$

In order to solve Eq. (8), by using state space, Eq. (12) will be as follows:

$$\begin{cases} \{y_1\} = \{\dot{q}\} \\ \{y_2\} = \{q\} \end{cases} \rightarrow \{\dot{y}_2\} = \{y_1\} \tag{12}$$

By substituting Eq. (12) in Eq. (8) and by converting the equations to matrix form, Eq. (13) is calculated as follows:

$$\begin{bmatrix} [M] & [0] \\ [0] & [I] \end{bmatrix} \begin{bmatrix} \{\dot{y}_1\} \\ \{\dot{y}_2\} \end{bmatrix} + \begin{bmatrix} [0] & [K] \\ -[I] & [0] \end{bmatrix} \begin{bmatrix} \{y_1\} \\ \{y_2\} \end{bmatrix} = \begin{bmatrix} \{Q\} \\ \{0\} \end{bmatrix} \tag{13}$$

After some algebraic operations, Eq. (13) is changed to a standard form as follows:

$$\begin{bmatrix} \{\dot{y}_1\} \\ \{\dot{y}_2\} \end{bmatrix} = \begin{bmatrix} [M] & [0] \\ [0] & [I] \end{bmatrix}^{-1} \left\{ -\begin{bmatrix} [0] & [K] \\ -[I] & [0] \end{bmatrix} \begin{bmatrix} \{y_1\} \\ \{y_2\} \end{bmatrix} + \begin{bmatrix} \{Q\} \\ \{0\} \end{bmatrix} \right\} \tag{14}$$

It can be seen that, the latest equation is in the form of $\dot{y} = f(t,y)$ and it is suitable for solving the equations numerically by Runge-Kutta method.

### 2.4. Calculation of general electrical power

By knowing the displacement function of the beam which is affected by the wave force, the electrical voltage and charge which are generated by piezoelectric patches in time t, is described as Eqs. (24) and (25) [17]:

$$Q_g^{pp}(t) = -e_{31} b \left( \frac{h + h_1}{2} \right) \times \left( \left. \frac{\partial w(z,t)}{\partial z} \right|_{z=-d+(pp.a)} - \left. \frac{\partial w(z,t)}{\partial z} \right|_{z=-d+((pp-1).a)} \right) \tag{15}$$

$$V_g^{pp}(t) = \frac{Q_g^{pp}(t)}{c_v} = -e_{31} \left( \frac{h + h_1}{2 c_v'} \right) \times \left( \left. \frac{\partial w(z,t)}{\partial z} \right|_{z=-d+(pp.a)} - \left. \frac{\partial w(z,t)}{\partial z} \right|_{z=-d+((pp-1).a)} \right) \tag{16}$$

where $e_{31}$, $h_1$, $c_v'$, $a$ and $N(1 \leq pp \leq N')$ are piezoelectric coefficient, thickness, electrical capacity per unit weight, length and number of the peizoelectric patches respectively. The generated output power in time $t$, can be defined as:

$$Pe(t) = \sum_{pp=1}^{N} \frac{dQ_g^{pp}(t)}{dt} V_g^{pp}(t) \qquad (17)$$

In addition, the average value of output power is calculated from Eq. (18), where $T$ is the total time.

$$Pe^{rms} = \sqrt{\frac{1}{T} \int_0^T [Pe(t)]^2 dt} \qquad (18)$$

## 3. Statistical analysis

In this section, the confidence interval which is used to calculate the error boundaries of intervals and the reliability and violation criteria are studied.

### 3.1. Confidence interval

In order to estimate the error boundaries of intervals including the actual values of the population parameter, the confidence interval is used [16]. For this purpose, it is considered that $X_1, X_2, ..., X_n$ are the random samples and $\theta$ is assumed to be an unknown parameter. Interval $(L,U)$ is the confidence interval for $\theta$ and is calculated by considering $X_1, X_2, ..., X_n$ before sampling. The mentioned interval includes an unknown actual value of $\theta$ with a certain probability. This probability is shown as "$1 - \alpha$" which is equal to 0.9. 0.95 or 0.99 most of the time. In other words, "$1 - \alpha$" is considered as a certain probability. Additionally, $L$ and $U$ are assumed as functions of $X_1, X_2, ..., X_n$, value of $1- \alpha$ is obtained as follows:

$$1 - \alpha = P[L < \theta < U] \qquad (19)$$

$(L, U)$ is the confidence interval of the coefficient 100% $(1 - \alpha)$, whereas $(1 - \alpha)$ is the confidence level of the population parameter. In order to clarify these concepts, confidence interval for the mean $\mu$ of data is considered when the sample size is large and the standard derivation is known. In next part, $\sigma$ is assumed to be unknown and this will result a more realistic formulation of the problem. Basis for description and expansion of confidence intervals is provided by possible form for the mean value of the sample *(X)* which is based on the normal distribution. According to the central limit theorem, the distribution of the $\bar{X}$ can be considered as $N(\mu, \frac{\sigma}{\sqrt{q}})$, in which, $\frac{\sigma}{\sqrt{q}}$ has a defined value. This distribution is a good approximation for a large sampling of the non-normal populations. But, whenever the population distribution is normal, the mentioned distribution holds for all values of $q$. Therefore, possible forms for the non-normal and normal populations are confirmed approximately and precisely, respectively. Generally, when $q$ has large value and the parameter $\sigma$ is defined, interval confidence of 100% (1- $\alpha$) for $\mu$ is obtained as:

$$(\bar{X} - Z_{\alpha/2} \frac{\sigma}{\sqrt{q}}, \bar{X} + Z_{\alpha/2} \frac{\sigma}{\sqrt{q}}) \tag{20}$$

where $Z_{\alpha/2}$ is the area of the right side of the standard normal distribution which is equal to $\alpha/2$. The mentioned quantities can be obtained from statistical table in [16].

### 3.2. The confidence interval based on large sample of μ and unknown value of σ

In the latest section, the basic concepts of confidence interval were defined and now, more realistic condition is considered in this section. In which, the standard deviation of the population is considered to be unknown. When the sample size of $\mu$ is large, the Eq. (20) is still correct. But, the interval cannot be acquired from the sample data since $\sigma$ is an unknown parameter in this section. Thus, it cannot be used as a confidence interval. Replacing $\sigma$ by its estimator (S) will not have a significant effect on the possible form of the Eq. (20), since $q$ is assumed to be large. Totally, when $q$ is large and standard deviation of the population is unknown, confidence interval of 100% (1- $\alpha$) for $\mu$ can be acquired as:

$$(\bar{X} - Z_{\alpha/2} \frac{S}{\sqrt{q}}, \bar{X} + Z_{\alpha/2} \frac{S}{\sqrt{q}}) \tag{21}$$

where, S is the standard deviation of the sample. The main assumption for population distribution is that the $\sigma$ has a predefined value.

### 3.3. Reliability and violation criteria

In this section, reliability of the system is studied. Reliability investigates the probability of failure based on the limit state function. It should be noted that, reliability is not limited to calculation of failure probability. Investigating properties of various statistical data such as: probability distribution functions and interval confidence are important in studying reliability. When a structure operates more than the expected limitation, the structure will lose the desirable function. The expected limitation is limit-state function. Therefore, system is considered in non-reliable condition, whenever there is probability of failure and violation from the limit-state. In other words, violation from a limitation for the system is considered as failure of the system. Limit-state is divided to two types [18]:

1- Failure of the structure

2- Disorder in normal operation

Generally, limit state reveals safety margin between load and strength. Equation related to limit state function and probability of failure are obtained as follows:

$$M = R - S \tag{22}$$

$$P_f = P(R \leq S) \tag{23}$$

where $R$ is the strength of the structure and $S$ is the load which is applied to the system. For a specific case, in which, $R$ and $S$ are normally distributed and are uncorrelated, limit state function will have normal distribution. Probability of failure is obtained by Eq. (24) as follows:

$$P(R - S \leq 0) = P(M \leq 0) \tag{24}$$

where $M$ is distributed normally and the mean value for this parameter is $\mu_M = \mu_R - \mu_S$ and standard deviation is $\sigma_M = \sqrt{\sigma_R^2 + \sigma_S^2}$. Probability of failure is acquired by normal distribution function as follows:

$$P_f(\frac{-\mu_M}{\sigma_M}) = \Phi(-\beta) \tag{25}$$

where $\beta = \dfrac{\mu_M}{\sigma_M}$ is the safety index and $\Phi$ is standard cumulative distribution function. Geometric description of the safety index is demonstrated in Fig.2. Shaded area of this figure shows the probability failure.

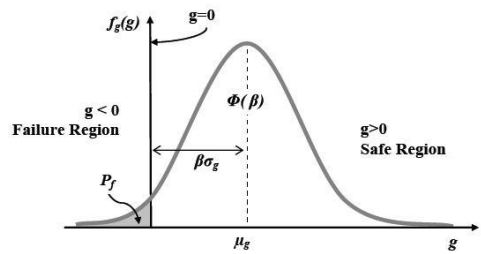

**Fig. 2** probability density for limit state

## 4. Simulation and results

Properties of the beam and characteristics of the waves which are used in simulation are summarized in Tables 1 and 2 respectively. By considering these properties, equations of the energy harvester system are solved by programming in Matlab and so, vibration response is determined. Subsequently, the output power produced from energy harvester is calculated.

**Table.1** properties of the beam

| Parameter name | Symbol | Value |
|---|---|---|
| beam length | $L(\text{m})$ | 3 |
| inertial coefficient of beam | $c_M$ | 1.7 |
| drag coefficient of beam | $c_D$ | 0.8 |
| Inertial coefficient of added mass | $c_m$ | 1 |
| Density of beam | $\rho(\text{kg}/\text{m}^2)$ | 7500 |
| Young module of beam | $E(\text{GPa})$ | 78 |
| Electrical capacity of patches | $c_v(\text{nF})$ | 0.75 |
| Piezoelectric constant | $e_{31}(\text{c}/\text{m}^2)$ | -2.8 |

**Table.2** properties of sea

| Parameter name | Symbol | Value |
|---|---|---|
| Sea depth | $d(\text{m})$ | 3 |
| Significant wave height | $H(\text{m})$ | 2 |
| Spectral peak period | $T(\text{s})$ | 15 |
| Sea water density | $\rho_w(\text{kg}/\text{m}^2)$ | 1025 |

By considering $f_{\max} = 63.52 Hz$ which is assumed to derive the maximum frequency of the system according to Nyquist theory $\Delta t = \dfrac{1}{2 f_{\max}}$. Number of discrete time data, $N = \dfrac{T}{\Delta t}$, and significant wave height $H_s = 2m$, the JONSWAP wave shape is illustrated in Fig.3.

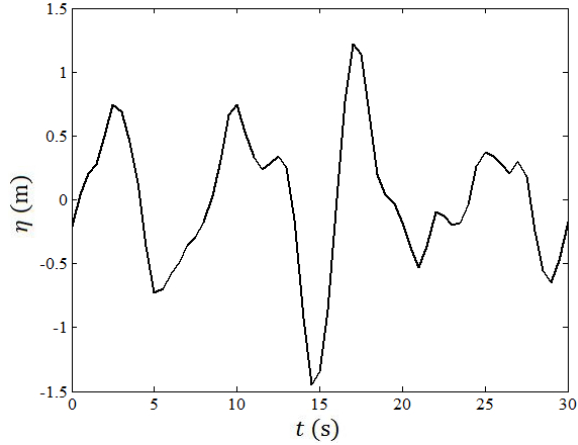

**Fig. 3** Irregular JONSWAP wave elevation

JONSWAP wave shape spectrum is compared with the one, which is obtained by Fourier transform of the signal, and is illustrated in Fig.4. It can be seen that, dominant frequency ($f_{peak} = \dfrac{1.2568}{2\pi\sqrt{H_s}}$) is visible in both situations.

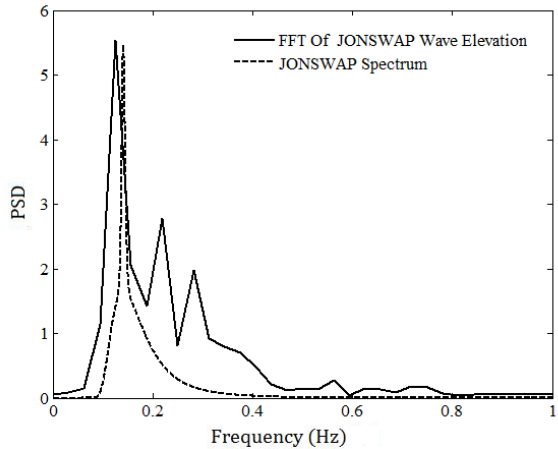

**Fig. 4** Comparison of JONSWAP wave spectrum with Fourier transform of JONSWAP wave elevation

Generated electrical power of the energy scavenging system by consideration of JONSWAP wave theory has not been studied yet. Therefore, in order to evaluate the system, dominant frequency of the system can be obtained by substituting *N=3812* and *H$_s$=2m* in $f_{peak} = \dfrac{1.2568}{2\pi\sqrt{H_s}}$. So the dominant frequency will be equal to 0.14Hz and it is compared to the one which is obtained by vibrational frequency response. Additionally, vibrational response of the beam in time domain is determined and subsequently by Fourier transformation of the response in time domain, the vibrational response in frequency domain is obtained. The responses in time and frequency domains are shown in Figs 5 and 6 respectively.

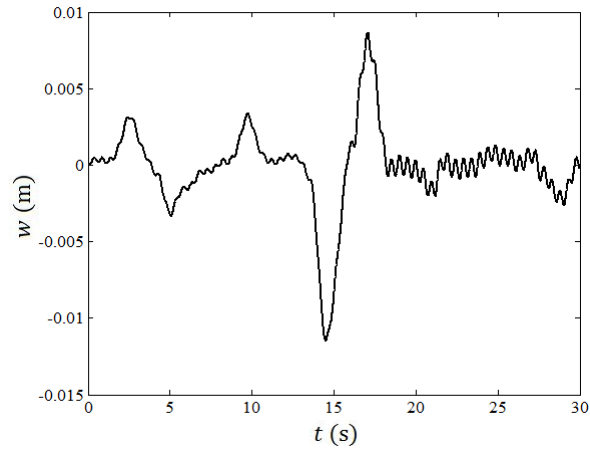

**Fig. 5** Free end displacement of energy scavenging system

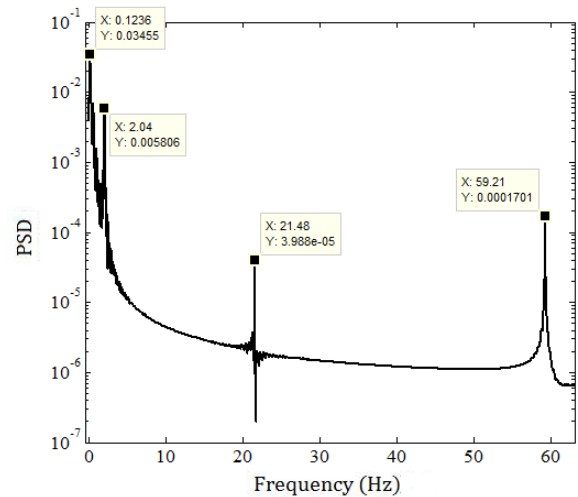

**Fig. 6** Fourier transform of free end displacement of energy scavenging system

As it can be seen from Fig.6, the appeared frequencies are equal to wave frequency and natural frequencies of the energy scavenging system. It should be noted that, natural frequencies of the system which are the roots of the character equation, is equal to 2.03, 20.35 and 63.52 Hz. It can be seen from Fig.6, natural frequencies of the beam and JONSWAP wave frequency is near to these values and this proves that the simulation of the system by consideration of JONSWAP wave theory is evaluated.

Now, uncertainties in length and thickness of piezoelectric patches with a nominal value of 0.1 and 0.001 meters are considered. Upper and lower limit of these parameters are assumed to be ±5% of the nominal value. Moreover, uncertainties are applied to the system by uniform distribution. Then, the RMS generated power as a function of beam and wave properties with consideration of JONSWAP wave theory is illustrated in Figs. 7 to 15. Furthermore, the average value of upper and lower limit of each figure is represented in Table 3.

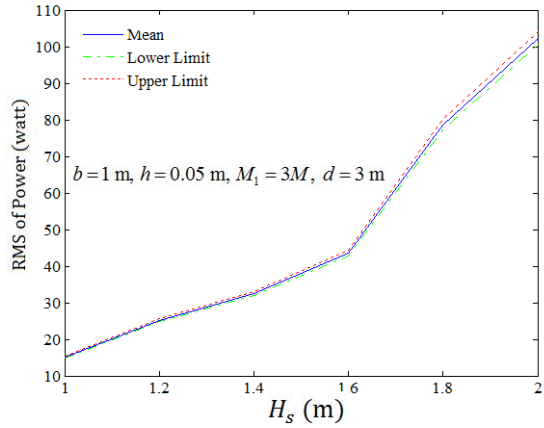

**Fig. 7** Generated electrical power versus significant wave height by consideration of uncertainty in length and thickness of patches

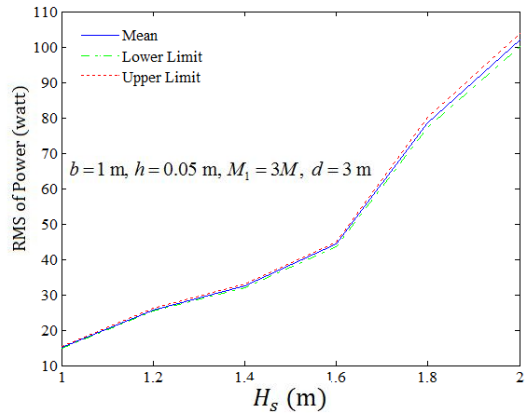

**Fig. 8** Generated electrical power versus significant wave height by consideration of uncertainty only in length of patches

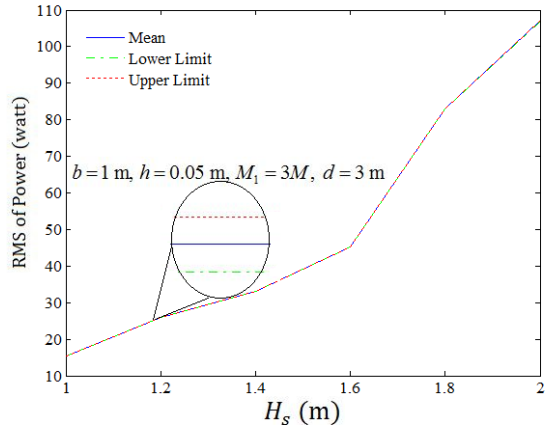

**Fig. 9** Generated electrical power versus significant wave height by consideration of uncertainty only in thickness of patches

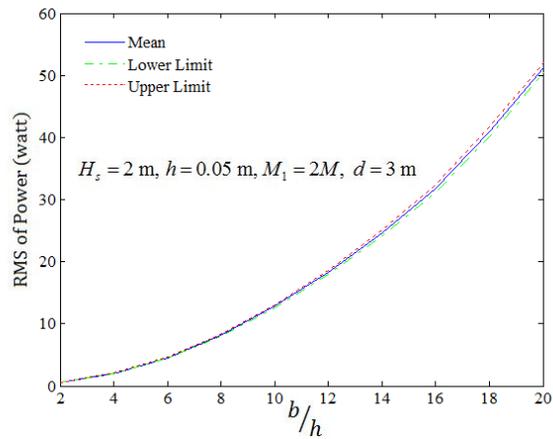

**Fig. 10** Generated electrical power versus width to thickness of the beam by consideration of uncertainty in length and thickness of patches

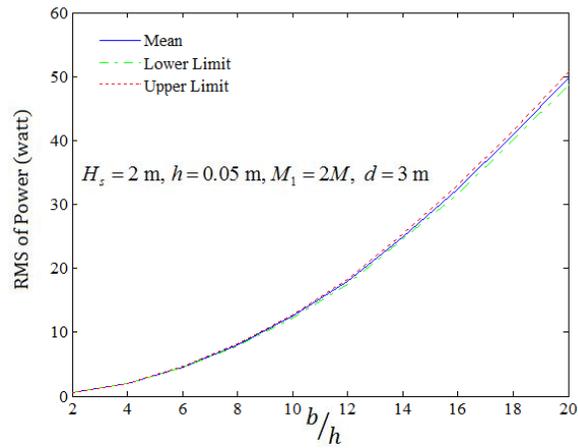

**Fig. 11** Generated electrical power versus width to thickness of the beam by consideration of uncertainty only in length

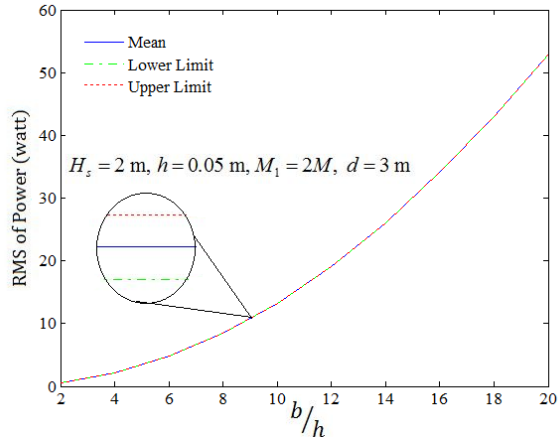

**Fig. 12** Generated electrical power versus width to thickness of the beam by consideration of uncertainty only in thickness of patches

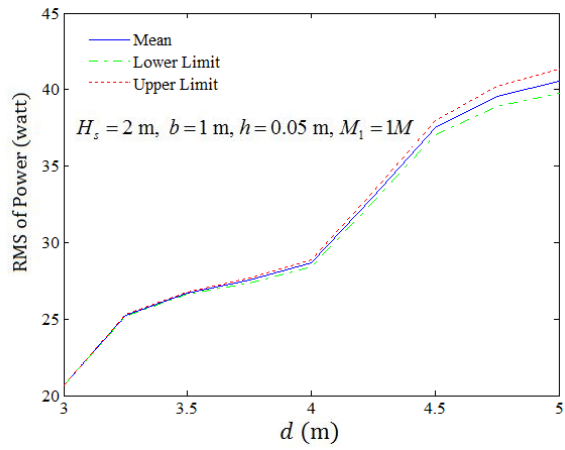

**Fig. 13** Generated electrical power versus sea depth by consideration of uncertainty in length and thickness of patches

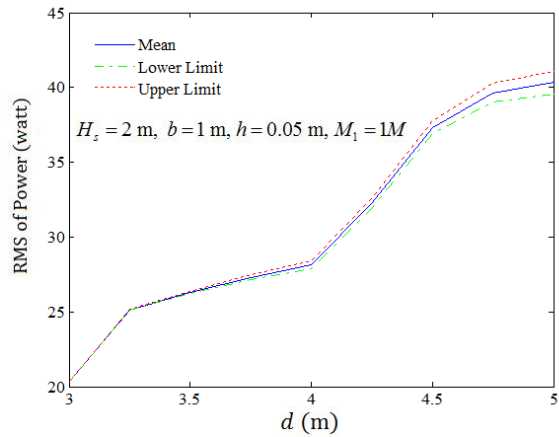

**Fig. 14** Generated electrical power versus sea depth by consideration of uncertainty only in length of patches

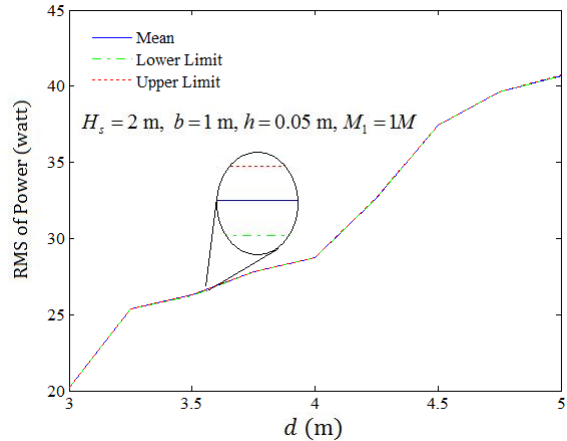

**Fig. 15** Generated electrical power versus sea depth by consideration of uncertainty only in thickness of patches

Table.3 the average value of upper and lower limit

| Figure | Uncertainty in length and thickness of patches | Uncertainty only in length of patches | Uncertainty only in thickness of patches |
|---|---|---|---|
| Generated power vs. significant wave height | 1.6358 | 1.7082 | 0.0190 |
| Generated power vs. width to thickness of the beam | 0.7143 | 0.7313 | 0.0072 |
| Generated power vs. sea depth | 0.6052 | 0.5967 | 0.0062 |

Generated electrical power versus significant wave height for upper limit, lower limit and for mean value is illustrated in Figs. 7 to 9. As it can be seen from these figures, by increasing the significant wave height, the generated electrical power increases as expected. Moreover, Figs 10 to 12 demonstrates the generated electrical power versus width to thickness of the beam for upper limit, lower limit and mean value. As it can be realized from these figures, increasing the width of the beam results growth in generated electrical power. Also, the mentioned electrical power versus sea depth which is considered equal to length of beam, for upper limit, lower limit and for mean value is illustrated in Figs. 13 to 15. It can be noticed that, by increasing the length of the beam, bending moment resulted from sea waves increases and consequently the generated electrical power increases. According to Table.3, generated electrical power is less sensitive to thickness of the patches in comparison with the patches length. When there is uncertainty only in thickness of the patches, upper limit, lower limit and mean value are more coincident. Though, whenever there is uncertainty in length and thickness of the patches simultaneously or only in length of the patches, the distances between the mentioned values are more. In other words, the generated electrical power is more sensitive to length of the patches. It can also be seen from these figures, augmentation of generated electrical power, and results incensement in uncertainties. According to the direct relationship between generated electrical power and generated voltage from piezoelectric patches, the violation criteria is considered such that, if the average value of generated electrical power is more than half of the maximum power in each run, failure is occurred.

In addition, the percentage of probability of failure which is resulted by illegal voltage, versus significant wave height, width to thickness of the beam and sea depth is studied and illustrated in Figs. 16 to 18. It can be noticed that, whenever the generated power increases, the probability of failure increases.

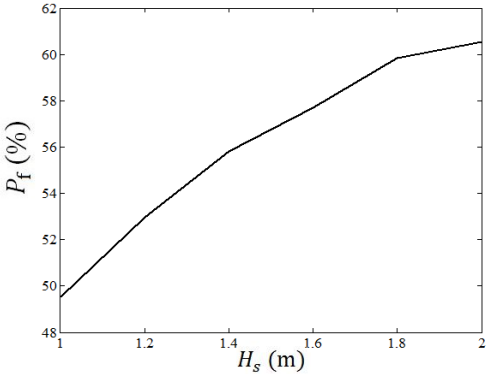

**Fig.16** Probability of failure versus significant wave height

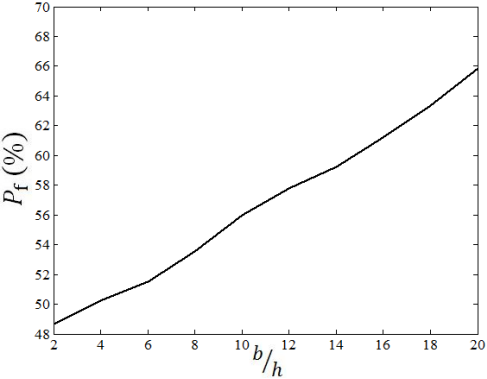

**Fig. 17** Probability of failure versus width to thickness of the beam

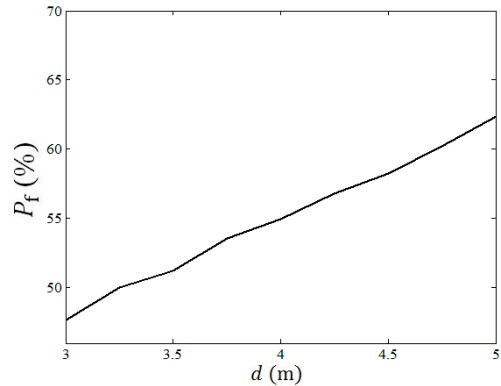

**Fig. 18** Probability of failure versus sea depth

## 5. Conclusion

Conversion of mechanical energies of sea waves into electrical energy by means of piezoelectric materials is considered as one of the most recent methods for powering low-power electronic devices at sea. The implemented energy harvesters are vibrated due to sea waves and as a result of vibration, dynamical strain is produced in the system and subsequently electrical power is generated. In this paper, simulation of energy harvesting system which is subjected to JONSWAP wave theory is investigated and the generated power is studied. Numerical methods are used for deriving the vibrational response. In the case of evaluating the modeling of the system, Fourier transform method is used and the resulted frequencies are compared with the natural frequencies which are determined by roots of the character equation and also with the wave frequency. After modeling the system and determining the equations, generated electrical power value for various parameters of the beam and sea is studied and also by consideration of uncertainties in length and thickness of piezoelectric patches, the mentioned value is investigated. In addition, lower limit, upper limit and mean value of the generated electrical power is defined. It is concluded that, increasing the significant wave height, width and length of the beam results an increase in the generated electrical power. Also, it is shown that, the generated electrical power is less sensitive to thickness of the patches. Furthermore, probability of failure is studied and it is noticed that, by increasing the power, the probability of failure increases.